\documentstyle[aps,preprint]{revtex}
\begin{document}
\author{Remo Garattini}
\address{Facolt\`a di Ingegneria, Universit\`a di Bergamo\\
Viale Marconi, 5, 24044 Dalmine (Bergamo) Italy\\
e-mail: Garattini@mi.infn.it}
\title{Vacuum Energy in Ultralocal metrics for TT tensors with\\
Gaussian Wave Functionals}
\date{July-1995}
\maketitle

\begin{abstract}
We calculate, in a class of Gauge invariant functionals, by variational
methods, the difference of vacuum energy between two different backgrounds:
Schwarzschild and Flat Space. We perform this evaluation in an Hamiltonian
formulation of Quantum Gravity by standard ''$3+1$'' decomposition. After
the decomposition the scalar curvature is expanded to second order with
respect to the Schwarzschild metric. We evaluate this energy difference in
momentum space, in the lowest possible state (regardless of any negative
mode). We find a singular behaviour in the UV-limit, due to the presence of
the horizon when $r=2m.$ When $r>2m$ this singular behaviour disappears,
which is in agreement with various other models presented in the literature.
\end{abstract}

\section{Introduction}

The problem of computing quantum corrections to a classical energy in a
complicated theory such as Einstein gravity, can be approached by performing
an analysis of the thermodynamical quantities that characterize the system
under consideration. This analysis can be done by means of the computation
of the free energy of the system at a given volume and temperature. Defining
the Euclidean action as 
\begin{equation}
\label{aa1}\hat I\left[ g\right] =-\frac 1{16\pi G}\int_{{\cal M}}d_{}^4x%
\sqrt{g}R\left( g\right) -\frac 1{8\pi G}\int_{{\cal \partial M}}d_{}^3x%
\sqrt{h}K_i^i, 
\end{equation}
where $R$ is the Ricci scalar of the metric $g_{\mu \nu }$ and $K_i^i$ is
the trace of the second fundamental form, one can compute quantum
corrections to the Euclidean action, of a fixed background geometry with the
appropriate boundary conditions. Since we wish to study quantum
fluctuactions with respect to the Schwarzschild geometry, there are two
types of boundary conditions that are related to the background under
consideration:

\begin{description}
\item[a)]  Asymptotically Euclidean (AE),

\item[b)]  Asymptotically Flat (AF).
\end{description}

An AE background metric is one in which the metric approaches the flat
metric on $R^4$ outside some compact set. The boundary at infinity is
topologically $S^3$. An AF background metric is one in which the metric
approaches the flat metric $R^3\times S^1$ outside some compact set. The
boundary of infinity is topologically $S^2\times S^1$. Anyway, we could
consider a different point of view based on the Hamiltonian approach. In
this framework one is able to deal with three dimensional fields
configurations separated out by the time variable. The advantage of the
Hamiltonian framework is that one can manage from the beginning with energy
fields configurations which give directly the quantum corrections to the
classical term. To do this, the first step is the separation of the three
dimensional space from the time by means of the ${\cal ADM}$ variables $\cite
{ADM}$. In terms of these variables the line element becomes 
$$
ds_{}^2=g_{\mu \nu }\left( x\right) dx^\mu dx^\nu =-N_{}^2\left( dx^0\right)
^2+g_{ij}\left( N^idx^0+dx^i\right) \left( N^jdx^0+dx^j\right) 
$$
\begin{equation}
\label{aa2}=\left( -N^2+N_iN^i\right) \left( dx^0\right)
^2+2N_jdx^0dx^j+g_{ij}dx^idx^j. 
\end{equation}
$N$ is called the {\it lapse function}, while $N_i$ the {\it shift function}%
. The associated matrix representation of $g_{\mu \nu }$ is

\begin{equation}
\label{aa3}g_{\mu \nu }=\left( 
\begin{array}{cc}
-N^2+N_iN^i & N_j \\ 
N_i & g_{ij} 
\end{array}
\right) , 
\end{equation}
with the inverse given by 
\begin{equation}
\label{aa3aa}g_{}^{\mu \nu }=\left( 
\begin{array}{cc}
-\frac 1{N^2} & \frac{N_{}^j}{N^2} \\ \frac{N^i}{N^2} & g_{ij}-\frac{N_iN^i}{%
N^2} 
\end{array}
\right) . 
\end{equation}

Roman indices will be raised and lowered by the induced metric on the three
surface $x^0$. In terms of the ${\cal ADM}$ variables, the initial action
can be written as a sum of a ``{\it kinetic}'' and a ``{\it potential}''
term 
\begin{equation}
\label{aa4}I=\frac 1{16\pi G}\int_{}dx_{}^0N\int dx_{}^3\text{ }^{\left(
3\right) }\sqrt{g}\left\{ \left( K_{ij}K^{ij}-K^2\right) +\text{ }^{\left(
3\right) }R\right\} ,
\end{equation}
where $K_{ij}^{}=\frac 1{2N}\left( N_{i|j}+N_{j|i}-g_{ij,0}\right) $ is
called the second fundamental form and ``%
%TCIMACRO{\TEXTsymbol{\vert}}
%BeginExpansion
\mbox{$\vert$}
%EndExpansion
'' means covariant differentiation with respect to the $3D$ gravitational
background, $K$ is the trace of the second fundamental form, $^3R$ is the
scalar curvature in $3D$, $^{\left( 3\right) }\sqrt{g}$ is the invariant of
the metric in $3D$. In this form the time derivative is isolated and it is
possible the computation of the conjugate momentum to $g_{ij}$, that is 
\begin{equation}
\label{aa5}\pi ^{ij}=\frac{\delta I}{\delta \dot g_{ij}}=\left( -K^{ij}+%
\text{ }^{\left( 3\right) }g^{ij}K\right) \frac{\sqrt{g}}{16\pi G}.
\end{equation}
By a Legendre transformation we calculate the Hamiltonian 
$$
H=\int d^3x\left\{ \left[ \pi ^{ij}\dot g_{ij}-\frac 1{16\pi G}\left[ \left(
K_{ij}K^{ij}-K^2\right) +\text{ }^{\left( 3\right) }R\right] N\text{ }%
^{\left( 3\right) }\sqrt{g}\right] \right\} = 
$$
\begin{equation}
\label{aa6}%
%TCIMACRO{\dint }
%BeginExpansion
\displaystyle \int 
%EndExpansion
d^3x\left\{ N\left[ \frac{16\pi G}{^{\left( 3\right) }\sqrt{g}}\left( \pi
_{ij}\pi ^{ij}-\frac \pi 2^2\right) -\text{ }^{\left( 3\right) }\sqrt{g}%
\frac{^{\left( 3\right) }R}{16\pi G}\right] +N_i\left( 2\pi
_{|j}^{ij}\right) \right\} .
\end{equation}
The first term of $\left( \ref{aa6}\right) $ has a quadratic structure in
the momenta, suggesting, as a first approximation that we could compute
quantum corrections to the energy expanding $^3R$ in terms of the quantum
field fluctuations with respect to a given background, e.g. the
Schwarzschild background. Since we wish to understand the pure gravitational
vacuum, we neglect the matter fields and since in our approach only the
spatial part of the background comes into play our background is of the
wormhole type \cite{MTW}. Although the energy computation at quantum level
is very unclear because of the constraint coming from the lapse function we
will adopt the Hamiltonian approach the same and to this purpose a simple
framework will be shown in section \ref{p2a}. The rest of the paper is
structured as follows, in section \ref{p2} we analyze the orthogonal
decomposition of the Hamiltonian both in tangent and co-tangent space, in
section \ref{p3} we define the gaussian wave functional for gravity in
analogy with non-abelian gauge theories, in section \ref{p4} we give some of
the basic rules to perform the functional integration and we define the
Hamiltonian approximated up to second order, in section \ref{p5}, we analyze
the spin-2 operator or the operator acting on transverse traceless tensors,
only for positive values of $E^2$. We summarize and conclude in section \ref
{p6}.

\section{ The Hamiltonian on the slice}

\label{p2a}

After the introduction of ${\cal ADM}$

variables, we recall that the Hamiltonian is: 
\begin{equation}
\label{a1}
\begin{array}{c}
H=\int d^3x(N{\cal H+}N_i{\cal H}^i) 
\end{array}
\end{equation}
where 
\begin{equation}
\label{a2}{\cal H}{\bf =}G_{ijkl}\pi ^{ij}\pi ^{kl}\left( \frac{l_p^2}{\sqrt{%
g}}\right) -\left( \frac{\sqrt{g}}{l_p^2}\right) \text{ }^{\left( 3\right)
}R\ \text{ (Super Hamiltonian)} 
\end{equation}
and

\begin{equation}
\label{a3}{\cal H}^i=-2\pi _{|j}^{ij}\ \text{ (Super Momentum).} 
\end{equation}
In $\left( \ref{a3}\right) $ the derivative is covariant with respect to the 
$3D$ background field, $l_p^2$ is the usual Planck mass, and $G_{ijkl}$ is
the Wheeler-DeWitt (WDW) metric. If we look at $N$ and $N_i$ as fundamental
objects describing the correct variables, by variational principles we
obtain the usual constraint equations, that is

\begin{equation}
\label{a3aaa}
\begin{array}{c}
{\cal H}\text{ }=0\text{, }{\cal H}^i\text{ }=0\text{ Classical} \\ {\cal H}%
\Psi \text{ }=0\text{, }{\cal H}^i\text{ }\Psi =0\text{ Quantum } \\  
\end{array}
\end{equation}
The usual interpretation of these equations is that they represent
constraints on the initial value problem or in other words they represent
gauge invariance with respect to time and gauge transformations.
Nevertheless we have a chance to define and computing energy if we restrict
on a given hypersurface, fixing the lapse function to a constant. Such a
gauge choice is the most appropriate for wormhole configurations of the
background geometry \cite{Garay} and in particular for the Schwarzschild
wormhole. By rescaling time intervals, we obtain\footnote{%
A different treatment, but close to our present approach, is based on the
separation between dynamical variables and embedding parameters and can be
found in Ref. \cite{Miller}} 
\begin{equation}
\label{a4}N=1. 
\end{equation}
Actually, this choice is compatible with the suspension constraint that one
can adopt in quantum cosmology to obtain a Schr\"odinger-like equation,
provided at the end of the calculation one assures that the gauge invariance
is restored \cite{Garay}\cite{J.J. Halliwell}. Then the Hamiltonian in the
time-like gauge is

\begin{equation}
\label{a5}H=\int d_{}^3x{\cal H}\text{ }=\int d_{}^3x\left[ G_{ijkl}\pi
^{ij}\pi ^{kl}\left( \frac{l_p^2}{\sqrt{g}}\right) -\left( \frac{\sqrt{g}}{%
l_p^2}\right) \text{ }^{\left( 3\right) }R\right] 
\end{equation}
Since $\left( \ref{a5}\right) $ is valid on a ``{\it fixed}'' hypersurface,
to recover the original equation, i.e. $\left( \ref{a1}\right) $, the
correct procedure to perform will be a summation over all possible lapses;
this means that the constraint $\left( \ref{a3aaa}\right) $ (at classical or
at quantum level) will be restored after this summation.

\section{Ultralocal Metrics as a tool for decomposing tensor fields}

\label{p2}

Instead of performing calculations in the usual WDW metric we will use a
one-parameter family of supermetrics to disentangle gauge modes from
physical deformations. For this reason we require an orthogonal
decomposition for both $\pi _{ij\text{ }}^{}$and $h_{ij}^{}$, that is we
need a metric on the space of deformations, i.e. a quadratic form on the
tangent space at h. The condition of ultralocality, where $G_{ijkl}$ locally
depends on $g_{ij}^{}$ but not on its derivatives, could be taken as a good
condition for the functional measure, explicitly:

\begin{equation}
\label{a6}\left\langle h,k\right\rangle :=\int_{{\cal M}}^{}\sqrt{g}G_\alpha
^{ijkl}h_{ij}^{}\left( x\right) k_{kl}^{}\left( x\right) d_{}^3x, 
\end{equation}

$$
\text{where} 
$$

\begin{equation}
\label{a7}
\begin{array}{c}
G_\alpha ^{ijkl}=(g_{}^{ik}g_{}^{jl}+g_{}^{il}g_{}^{jk}-2\alpha
g_{}^{ij}g_{}^{kl}).
\end{array}
\end{equation}
The WDW metric, introduced in $\left( \ref{a2}\right) $, is just $\left( \ref
{a7}\right) $ with $\alpha =1$. The ``inverse'' metric is defined on
co-tangent space and it assumes the form

\begin{equation}
\label{a8}\left\langle p,q\right\rangle :=\int_{{\cal M}}^{}\sqrt{g}%
G_{ijkl}^\beta p_{}^{ij}\left( x\right) q_{}^{kl}\left( x\right) d_{}^3x, 
\end{equation}

$$
\text{where} 
$$

\begin{equation}
\label{a9}
\begin{array}{c}
G_{ijkl}^\beta =(g_{ik}^{}g_{jl}^{}+g_{il}^{}g_{jk}^{}-2\beta
g_{ij}^{}g_{kl}^{}). \\  
\end{array}
\end{equation}
with $\alpha +\beta =3\alpha \beta $, so that

\begin{equation}
\label{a10}G_\beta ^{ijnm}G_{nmkl}^\beta =\frac 12\left( \delta _k^i\delta
_l^j+\delta _l^i\delta _k^j\right) . 
\end{equation}
These are non-degenerate bilinear forms for $\alpha \neq \frac 1{3\text{ }}$%
, for $\alpha =\frac 1{3\text{ }}$ the metric is not invertible and becomes
a projector onto the tracefree subspace, while is positive definite for $%
\alpha <\frac 1{3\text{ }}$ and of mixed signature for $\alpha >\frac 1{3%
\text{ }}$ with infinitely many plus as well as minus signs.

We have now the desired decomposition on the tangent space of 3-metric
deformations $h_{ij}^{}$:

\begin{equation}
\label{a11}h_{ij}^{}=\frac 13hg_{ij}^{}+\left( L\xi \right)
_{ij}^{}+h_{ij}^{\bot }, 
\end{equation}

$$
\text{or, in matrix form,} 
$$

\begin{equation}
\label{a12}h=\frac 13hg+\left( RangeL\right) +\left( KerL^{\dagger }\right)
, 
\end{equation}
where the operator $L$ maps $\xi _i^{}$ into symmetric tracefree tensors,
according to\cite{MazMot}\cite{York}$,$

\begin{equation}
\label{a13}\left( L\xi \right) _{ij}^{}=\nabla _i^{}\xi _j^{}+\nabla
_j^{}\xi _i^{}-\frac 23g_{ij}^{}\left( \nabla \cdot \xi \right) . 
\end{equation}
Consequently, the inversion of the metric $\left( \ref{a8}\right) \left( 
\text{that is}\left( \ref{a9}\right) \right) ,$guarantees us the same
decomposition also in phase space (co-tangent space).

\section{The Gaussian Wave Functional}

\label{p3}

There are some reasons to introduce a gaussian wave functional for the
description of the vacuum state in gravity. Starting from the analogy
between nonabelian gauge theories and gravity we illustrate how gaussian
wave functional work in the former case. We define \cite{Kerman}

\begin{equation}
\label{b1}\Psi \left[ A_i^a\left( \overrightarrow{x}\right) \right] ={\cal N}%
\exp \left\{ -\frac 14\int d_{}^3xd_{}^3y\delta A_i^a\left( \overrightarrow{x%
}\right) G_{ij}^{-1ab}\left( \overrightarrow{x},\overrightarrow{y}\right)
\delta A_j^b\left( \overrightarrow{y}\right) \right\} 
\end{equation}
where ${\cal N}$ is a normalization factor and where 
\begin{equation}
\label{b2}\delta A_i^a\left( \overrightarrow{x}\right) =A_i^a\left( 
\overrightarrow{x}\right) -\overline{A_i^a}\left( \overrightarrow{x}\right)
. 
\end{equation}
In equation $\left( \ref{b2}\right) $, $\overline{A_i^a}\left( 
\overrightarrow{x}\right) $ is a background field which can be treated as a
variational parameter together to the function $G_{ij}^{ab}\left( 
\overrightarrow{x},\overrightarrow{y}\right) $in $\left( \ref{b1}\right) $.

From the definition in $\left( \ref{b1}\right) $ one finds the expectation
values 
\begin{equation}
\label{b3}
\begin{array}{c}
\left\langle \Psi |A_i^a\left( 
\overrightarrow{x}\right) |\Psi \right\rangle =\overline{A_i^a}\left( 
\overrightarrow{x}\right) , \\  \\ 
\left\langle \Psi |A_i^a\left( 
\overrightarrow{x}\right) A_j^b\left( \overrightarrow{y}\right) |\Psi
\right\rangle =\overline{A_i^a}\left( \overrightarrow{x}\right) \overline{%
A_j^b}\left( \overrightarrow{y}\right) +G_{ij}^{ab}\left( \overrightarrow{x},%
\overrightarrow{y}\right) , \\  \\ 
\left\langle \Psi |E_i^a\left( 
\overrightarrow{x}\right) |\Psi \right\rangle =0, \\  \\ 
\left\langle \Psi |E_i^a\left( 
\overrightarrow{x}\right) E_j^b\left( \overrightarrow{y}\right) |\Psi
\right\rangle =\frac 14G_{ij}^{-1ab}\left( \overrightarrow{x},%
\overrightarrow{y}\right) , \\  \\ 
\left\langle \Psi |B_i^a\left( \overrightarrow{x}\right) |\Psi \right\rangle
=\overline{B_i^a}\left( \overrightarrow{x}\right) +\frac 12\epsilon
_{ijk}^{}f^{abc}G_{ij}^{ab}\left( \overrightarrow{x},\overrightarrow{y}%
\right) ,
\end{array}
\end{equation}
where 
\begin{equation}
\label{b4}E_i^a\left( \overrightarrow{x}\right) =-i\frac \delta {\delta
A_i^a\left( \overrightarrow{x}\right) },
\end{equation}
and 
\begin{equation}
\label{b5}B_i^a\left( \overrightarrow{x}\right) =\epsilon _{ijk}\left\{
\nabla _jA_k^a\left( \overrightarrow{x}\right) +\frac 12f^{abc}A_i^a\left( 
\overrightarrow{x}\right) A_j^b\left( \overrightarrow{y}\right) \right\} ,
\end{equation}
$\epsilon _{ijk}$ is the usual anti-symmetric tensor and $f^{abc}$ are the
structure constants of the gauge group, for ex. $SU\left( N\right) $. After
having experienced how the apparatus works on nonabelian gauge theories, we
define a ``Vacuum Trial State'' for gravity, and for this purpose we recall
the orthogonal decomposition $\left( \ref{a10}\right) $ to look at the
essential structure of the inner product between three-geometries%
$$
\left\langle h,h\right\rangle :=\int_{{\cal M}}^{}\sqrt{g}G_\alpha
^{ijkl}h_{ij}^{}\left( x\right) h_{kl}^{}\left( x\right) d_{}^3x= 
$$
\begin{equation}
\label{b6}\int_{{\cal M}}^{}\sqrt{g}\left[ \left( \frac 13-\alpha \right)
h^2+\left( L\xi \right) ^{ij}\left( L\xi \right) _{ij}^{}+h_{}^{ij\bot
}h_{ij}^{\bot }\right] 
\end{equation}
Previous formula leads us towards the definition of the ``{\it possible}''
trial wave functional for the gravitational ground state 
\begin{equation}
\label{b7}\Psi _\alpha \left[ h_{ij}^{}\left( \overrightarrow{x}\right)
\right] ={\cal N}\exp \left\{ -\frac 1{4l_p^2}\left[ \left\langle
hK_{}^{-1}h\right\rangle _{x,y}^{\bot }+\left\langle \left( L\xi \right)
K_{}^{-1}\left( L\xi \right) \right\rangle _{x,y}^{\Vert }+\left\langle
hK_{}^{-1}h\right\rangle _{x,y}^{Trace}\right] \right\} ,
\end{equation}
or in other terms 
\begin{equation}
\label{b8}\Psi _\alpha \left[ h_{ij}^{}\left( \overrightarrow{x}\right)
\right] ={\cal N}\Psi _\alpha \left[ h_{ij}^{\bot }\left( \overrightarrow{x}%
\right) \right] \Psi _\alpha \left[ \left( L\xi \right) _{ij}^{}\right] \Psi
_\alpha \left[ \frac 13g_{ij}^{}h\left( \overrightarrow{x}\right) \right] .
\end{equation}
In $\left( \ref{b7}\right) $ and in $\left( \ref{b8}\right) ,$ $h_{ij}^{\bot
}$ is the tracefree-transverse part of the $3D$ quantum field, $\left( L\xi
\right) _{ij}^{}$ is the longitudinal part and finally $h$ is the trace part
of the same field. The dependence of the functional by $\alpha $ will not be
discussed in this paper.

In $\left( \ref{b7}\right) $, $\left\langle \cdot ,\cdot \right\rangle
_{x,y}^{}$ denotes space integration and $K_{}^{-1}$ is the inverse
propagator. The main reason for a similar ``{\it Ansatz}'' comes not only
from $\left( \ref{b6}\right) $ but even to the observation that the momenta
quadratic part of the Hamiltonian decouples in the same way. Even if we had
to give up to $\left( \ref{b7}\right) $ from the beginning, making a more
general ``{\it Ansatz}'' about the vacuum wave functional (and for more
general we mean eqn. $\left( \ref{b1}\right) $ ) one would discover that the
kinetic part decouples in these three terms. For completeness, we give the
analogous expectation values for TT tensors. The other components satisfy
the same rules 
\begin{equation}
\label{b9}
\begin{array}{c}
\left\langle \Psi |g_{ij}^{\bot }\left( 
\overrightarrow{x}\right) |\Psi \right\rangle =\bar g_{ij}^{\bot }\left( 
\overrightarrow{x}\right) , \\  \\ 
\left\langle \Psi |g_{ij}^{\bot }\left( 
\overrightarrow{x}\right) g_{kl}^{\bot }\left( \overrightarrow{y}\right)
|\Psi \right\rangle =\bar g_{ij}^{\bot }\left( \overrightarrow{x}\right) 
\bar g_{kl}^{\bot }\left( \overrightarrow{y}\right) +K_{ijkl}^{\bot }\left( 
\overrightarrow{x},\overrightarrow{y}\right) , \\  \\ 
\left\langle \Psi |\pi _{ij}^{\bot }\left( 
\overrightarrow{x}\right) |\Psi \right\rangle =0, \\  \\ 
\left\langle \Psi |\pi _{ij}^{\bot }\left( 
\overrightarrow{x}\right) \pi _{kl}^{\bot }\left( \overrightarrow{y}\right)
|\Psi \right\rangle =\frac 14K_{ijkl}^{-1}\left( \overrightarrow{x},%
\overrightarrow{y}\right) , \\ 
\end{array}
\end{equation}
where $\pi _{ij}^{\bot }=-i\frac \delta {\delta h_{ij^{}}^{\bot }\left(
x\right) }$ is the representation for the TT momentum.

\section{Energy Density Calculation in Schr\"odinger Representation}

\label{p4}

To calculate the energy density associated to the trial functional, we need
to know the action of some basic operators on $\Psi \left[ h_{ij}^{}\left( 
\overrightarrow{x}\right) \right] $. The action of the operator $h_{ij}^{}$
on $|\Psi \rangle =\Psi \left[ h_{ij}^{}\left( \overrightarrow{x}\right)
\right] $ is realized by 
\begin{equation}
\label{c1}h_{ij}^{}\left( x\right) |\Psi \rangle =h_{ij}^{}\left( x\right)
\Psi \left[ h_{ij}^{}\left( \overrightarrow{x}\right) \right] . 
\end{equation}
The action of the operator $\pi _{ij}^{}$ on $|\Psi \rangle $, in general, is

\begin{equation}
\label{c2}\pi _{ij}^{}\left( x\right) |\Psi \rangle =-i\frac \delta {\delta
h_{ij^{}}^{}\left( x\right) }\Psi \left[ h_{ij}^{}\left( \overrightarrow{x}%
\right) \right] . 
\end{equation}
The inner product is defined by the functional integration: 
\begin{equation}
\label{c3}\left\langle \Psi _1\mid \Psi _2\right\rangle =\int \left[ {\cal D}%
h_{ij}^{}\left( x\right) \right] \Psi _1^{*}\left\{ h_{ij}^{}\right\} \Psi
_2\left\{ h_{kl}^{}\right\} , 
\end{equation}
and the energy eigenstates satisfy the Schr\"odinger equation: 
\begin{equation}
\label{c4}\int d_{}^3x{\cal H}\left\{ -i\frac \delta {\delta
h_{ij^{}}^{}\left( x\right) },h_{ij}\left( x\right) \right\} \Psi \left\{
h_{ij}^{}\left( \overrightarrow{x}\right) \right\} =E\Psi \left\{
h_{ij}^{}\left( \overrightarrow{x}\right) \right\} , 
\end{equation}
where ${\cal H}\left\{ -i\frac \delta {\delta h_{ij^{}}^{}\left( x\right) }%
,h_{ij}\left( x\right) \right\} $ is the Hamiltonian density. Instead of
solving $\left( \ref{c4}\right) $, which is of course impossible, we can
formulate the same problem by means of a variational principle. We demand
that 
\begin{equation}
\label{c6}\frac{\left\langle \Psi _{}^{}\mid H\mid \Psi _{}\right\rangle }{%
\left\langle \Psi _{}^{}\mid \Psi _{}\right\rangle }=\frac{\int \left[ {\cal %
D}g_{ij}^{\bot }\left( x\right) \right] \int d_{}^3x\Psi _1^{*}\left\{
g_{ij}^{\bot }\right\} {\cal H}\Psi \left\{ g_{kl}^{\bot }\right\} }{\int
\left[ {\cal D}g_{ij}^{\bot }\left( x\right) \right] \mid \Psi \left\{
g_{ij}^{\bot }\right\} \mid _{}^2} 
\end{equation}
be stationary against arbitrary variations of $\Psi \left\{ h_{ij}^{}\left( 
\overrightarrow{x}\right) \right\} $. The form of $\left\langle \Psi
_{}^{}\mid H\mid \Psi _{}\right\rangle $ can be computed as follows. We
define normalized mean values by a straightforward modification of $\left( 
\ref{b9}\right) $, i.e. 
\begin{equation}
\label{c7}\bar g_{ij}^{\bot }\left( x\right) =\frac{\int \left[ {\cal D}%
g_{ij}^{\bot }\left( x\right) \right] \int d_{}^3xg_{ij}^{\bot }\left(
x\right) \mid \Psi \left\{ g_{ij}^{\bot }\left( x\right) \right\} \mid _{}^2%
}{\int \left[ {\cal D}g_{ij}^{\bot }\left( x\right) \right] \mid \Psi
\left\{ g_{ij}^{\bot }\right\} \mid _{}^2}, 
\end{equation}
\begin{equation}
\label{c8}\bar g_{ij}^{\bot }\left( x\right) \text{ }\bar g_{kl}^{\bot
}\left( x\right) +K_{ijkl^{}}^{\bot }\left( \overrightarrow{x},%
\overrightarrow{y}\right) =\frac{\int \left[ {\cal D}g_{ij}^{\bot }\left(
x\right) \right] \int d_{}^3xg_{ij}^{\bot }\left( x\right) g_{kl}^{\bot
}\left( y\right) \mid \Psi \left\{ g_{ij}^{\bot }\left( x\right) \right\}
\mid _{}^2}{\int \left[ {\cal D}g_{ij}^{\bot }\left( x\right) \right] \mid
\Psi \left\{ g_{ij}^{\bot }\right\} \mid _{}^2}. 
\end{equation}
It follows that%
$$
\int \left[ {\cal D}h_{ij}^{\bot }\left( x\right) \right] \left(
g_{ij}^{\bot }\left( x\right) -\bar g_{ij}^{\bot }\left( x\right) \right)
\mid \Psi \left\{ g_{ij}^{\bot }\left( x\right) \right\} \mid _{}^2=0 
$$
by translation invariance of the measure%
$$
\int \left[ {\cal D}h_{ij}^{\bot }\left( x\right) \right] h_{ij}^{\bot
}\left( x\right) \mid \Psi \left\{ g_{ij}^{\bot }\left( x\right) +\bar g%
_{ij}^{\bot }\left( x\right) \right\} \mid _{}^2=0 
$$
\begin{equation}
\label{c9}\Longrightarrow \int \left[ {\cal D}h_{ij}^{\bot }\left( x\right)
\right] h_{ij}^{\bot }\left( x\right) \mid \Psi \left\{ h_{ij}^{\bot }\left(
x\right) \right\} \mid _{}^2=0, 
\end{equation}
and $\left( \ref{c8}\right) $ becomes 
\begin{equation}
\label{c10}
\begin{array}{c}
\int \left[ 
{\cal D}h_{ij}^{\bot }\left( x\right) \right] \int d_{}^3xh_{ij}^{\bot
}\left( x\right) h_{kl}^{\bot }\left( y\right) \mid \Psi \left\{
h_{ij}^{\bot }\left( x\right) \right\} \mid _{}^2= \\  \\ 
K_{ijkl^{}}^{\bot }\left( \overrightarrow{x},\overrightarrow{y}\right) \int
\left[ {\cal D}h_{ij}^{\bot }\left( x\right) \right] \mid \Psi \left\{
h_{ij}^{\bot }\right\} \mid _{}^2. 
\end{array}
\end{equation}
Rather than applying the variational principle arbitrarily, the gaussian 
{\it Ansatz} is made, according to which in the beginning of this calculus
one has to replace previous general formulas with 
\begin{equation}
\label{c11}\Psi _\alpha \left[ h_{ij}^{}\left( \overrightarrow{x}\right)
\right] ={\cal N}\exp \left\{ -\frac 1{4l_p^2}\left\langle \left( g-%
\overline{g}\right) K_{}^{-1}\left( g-\overline{g}\right) \right\rangle
_{x,y}^{\bot }+\ldots \ldots \right\} . 
\end{equation}
With this choice and with formulas $\left( \ref{c9},\ref{c10}\right) $, the
one loop-like Hamiltonian can be written as 
\begin{equation}
\label{c12}H_{}^{\bot }=\frac 1{4l_p^2}\int_{{\cal M}}^{}d_{}^3x\sqrt{g}%
G_\alpha ^{ijkl}\left[ K_{}^{-1\bot }\left( x,x\right) _{ijkl}+\left(
\triangle _2^{}\right) _j^aK_{}^{\bot }\left( x,x\right) _{iakl}\right] 
\end{equation}
where the first term in square brackets comes from the kinetic part and the
second comes from the expansion of the $^3R$ up to second order in such a
way to obtain a quantum harmonic oscillator equation type. the Green
function $K_{}^{\bot }\left( x,x\right) _{iakl}$ can be represented as 
\begin{equation}
\label{ffi}K_{}^{\bot }\left( x,x\right) _{iakl}:=\sum_N\frac{h_{ia}^{\bot
}\left( x\right) h_{kl}^{\bot }\left( y\right) }{2\lambda _N\left( p\right) }%
, 
\end{equation}
where $h_{ia}^{\bot }\left( x\right) $ are the eigenfunctions of $\triangle
_{2j}^a$ and $\lambda _N\left( p\right) $ are infinite variational
parameters. In formula $\left( \ref{c12}\right) $ we have written the Spin-2
contribution to the energy density alone; expressions like $\left( \ref{c12}%
\right) $ exist for Spin-1 and Spin-0 terms of ${\cal H}$.

\section{The Spectrum of the Spin-2 Operator and the evaluation of the
Energy Density}

\label{p5}

The Spin-2 operator is defined by:

\begin{equation}
\label{ff1}\triangle _2^{}:=-\triangle +2Ric 
\end{equation}

$$
\text{or in components,} 
$$

\begin{equation}
\label{ff2}\left( \triangle _2^{}\right) _j^a:=-\triangle \delta
_j^{a_{}^{}}+2R_j^a 
\end{equation}
where $\triangle $ is the curved Laplacian (Laplace-Beltrami operator) on a
Schwarzschild background and

$R_{j\text{ }}^a$ is the mixed Ricci tensor whose components are:

\begin{equation}
\label{ff3}R_j^a=diag\left\{ \frac{-2m}{r_{}^3},\frac m{r_{}^3},\frac m{%
r_{}^3}\right\} , 
\end{equation}
where $2m=2MG$. This operator is similar to the Lichnerowicz operator
provided that we substitute the Riemann tensor with the Ricci tensor. In $%
\left( \ref{ff1}\right) $ or $\left( \ref{ff2}\right) $ Ricci tensor acts as
a potential on the space of TT tensors; for this reason we are led to study
the following eigenvalue equation

\begin{equation}
\label{ff4}\left( -\triangle \delta _j^{a_{}^{}}+2R_j^a\right)
h_a^i=E^2h_j^{i_{}^{}} 
\end{equation}
where $E^2$ is the eigenvalue of the corresponding equation. In doing so, we
follow Regge and Wheeler in analyzing the equation into modes of definite
frequency, angular momentum and parity. In this paper we are interested to
positive $E^2$ and low lying states with $L=M=0$, where $L$ is the quantum
number corresponding to the square of angular momentum and $M$ is the
quantum number corresponding to the projection of the angular momentum on
the z-axis. For $L=0$, Regge-Wheeler decomposition \cite{Regge} shows that
there are no odd-parity perturbations at all, therefore:

\begin{equation}
\label{ff5}h_{ij}^{even}=diag\left[ H\left( r\right) \left( 1-\frac{2m}r%
\right) _{}^{-1},r^2K\left( r\right) ,r_{}^2\sin _{}^2\vartheta K\left(
r\right) \right] Y_{00}^{}\left( \vartheta ,\phi \right) . 
\end{equation}
The representation $\left( \ref{ff5}\right) $ is very useful, because of the
decoupling of the components, in fact

\begin{equation}
\label{ff6}
\begin{array}{c}
-\triangle H\left( r\right) - 
\frac{4m}{r_{}^3}H\left( r\right) =E^2H\left( r\right) \\  \\ 
-\triangle K\left( r\right) + 
\frac{2m}{r_{}^3}K\left( r\right) =E^2K\left( r\right) \\  \\ 
-\triangle K\left( r\right) +\frac{2m}{r_{}^3}K\left( r\right) =E^2K\left(
r\right) 
\end{array}
\end{equation}
The Laplacian in this particular geometry can be written as

\begin{equation}
\label{ff7}\triangle =\left( 1-\frac{2m}r\right) \frac{d_{}^2}{dr_{}^2}%
+\left( \frac{2r-3m}{r_{}^2}\right) \frac d{dr}. 
\end{equation}
Defining reduced fields, such as:

\begin{equation}
\label{ff8}H\left( r\right) =\frac{h\left( r\right) }r;K\left( r\right) =%
\frac{k\left( r\right) }r, 
\end{equation}
and changing variables to

\begin{equation}
\label{ff9}x=2m\left\{ \sqrt{\frac r{2m}}\sqrt{\frac r{2m}-1}+\ln \left( 
\sqrt{\frac r{2m}}+\sqrt{\frac r{2m}-1}\right) \right\} , 
\end{equation}
the system $\left( \ref{ff6}\right) $ becomes

\begin{equation}
\label{ff10}
\begin{array}{c}
- 
\frac{d_{}^2}{dx_{}^2}h\left( x\right) -V\left( x\right) h\left( x\right)
=E^2h\left( x\right) \\  \\ 
- 
\frac{d_{}^2}{dx_{}^2}k\left( x\right) +V\left( x\right) k\left( x\right)
=E^2k\left( x\right) \\  \\ 
-\frac{d_{}^2}{dx_{}^2}k\left( x\right) +V\left( x\right) k\left( x\right)
=E^2k\left( x\right) 
\end{array}
\end{equation}
where 
\begin{equation}
\label{ff11}V\left( x\right) =\frac{3m}{r_{}^3} 
\end{equation}
We note that the new variable is such that 
\begin{equation}
\label{ff12}
\begin{array}{c}
x\simeq r 
\text{ }r\longrightarrow \infty \text{ }V\left( x\right) \longrightarrow 0
\\  \\ 
x\simeq 0\text{ }r\longrightarrow r_0\text{ }V\left( x\right)
\longrightarrow \frac{3m}{\left( r_0\right) _{}^3}=const, 
\end{array}
\end{equation}
where $r_0^{}$ is the wormhole radius, satisfying the condition $r_0^{}>2m$,
strictly. The solution of $\left( \ref{ff10}\right) $, in both cases (flat
and curved one) is a Bessel function and precisely the spherical Bessel
function of the first kind for the $L=0$ value of the angular momentum 
\begin{equation}
\label{ff13}j_0^{}\left( pr\right) =p\sqrt{\frac 2\pi }\frac{\sin \left(
pr\right) }{pr}=\sqrt{\frac 2\pi }\frac{\sin \left( pr\right) }r 
\end{equation}
The corresponding Green function for this problem will be 
\begin{equation}
\label{ff14}K\left( x,y\right) =\frac{j_0^{}\left( px\right) j_0^{}\left(
py\right) }{2\lambda }\cdot \frac 1{4\pi } 
\end{equation}
Substituting $\left( \ref{ff14}\right) $ in $\left( \ref{c12}\right) $ one
gets (after normalization in spin space and after a rescaling of the fields
in such a way to absorb $l_p^2$) 
\begin{equation}
\label{ff15}E\left( m,\lambda \right) =\frac V{2\pi ^2}\sum_{i=1}^2\int_0^%
\infty dpp_{}^2\left[ \lambda _i^{}\left( p\right) +\frac{E_i^2\left(
p,m\right) }{\lambda _i^{}\left( p\right) }\right] 
\end{equation}
where 
\begin{equation}
\label{ff16}E_{1,2}^2\left( p,m\right) =p_{}^2\mp \frac{3m}{r_0^3}, 
\end{equation}
$\lambda _i^{}\left( p\right) $ are variational parameters corresponding to
the eigenvalues for a (graviton) spin-2 particle in an external field and $V$
is the volume of the system.

By minimizing $\left( \ref{ff15}\right) $ with respect to $\lambda
_i^{}\left( p\right) $ one obtains $\overline{\lambda }_i^{}\left( p\right)
=\left[ E_i^2\left( p,m\right) \right] ^{\frac 12}$ and 
\begin{equation}
\label{ff17}E\left( m,\overline{\lambda }\right) =\frac V{2\pi ^2}%
\sum_{i=1}^2\int_0^\infty dp2\sqrt{E_i^2\left( p,m\right) }\text{ with }%
p_{}^2>\frac{3m}{r_0^3}
\end{equation}
The total energy in the presence of the background is 
\begin{equation}
\label{ff18}E\left( m\right) =\frac V{2\pi ^2}\frac 12\int_0^\infty
dpp_{}^2\left( \sqrt{p_{}^2-c_{}^2}+\sqrt{p_{}^2+c_{}^2}\right) \text{ where 
}c_{}^2=\frac{3m}{r_0^3}
\end{equation}
For flat space the calculation is essentially the same with the exception of 
$c_{}^2=0$. Therefore the equivalent of $\left( \ref{ff18}\right) $ in flat
space is 
\begin{equation}
\label{ff19}E\left( 0\right) =\frac V{2\pi ^2}\frac 12\int_0^\infty
dpp_{}^2\left( 2\sqrt{p^2}\right) 
\end{equation}
Now, we are in position to perform the energy difference between $\left( \ref
{ff18}\right) $and $\left( \ref{ff19}\right) $. $\Delta E\left( m\right) $
up to second order in perturbations is 
\begin{equation}
\label{ff20}\Delta E\left( m\right) =\frac V{2\pi ^2}\frac 12\int_0^\infty
dpp_{}^2\left[ \sqrt{p_{}^2-c_{}^2}+\sqrt{p_{}^2+c_{}^2}-2\sqrt{p^2}\right] 
\end{equation}
We want to evaluate the $UV$ behaviour of $\left( \ref{ff20}\right) $,
therefore%
$$
\begin{array}{c}
\Delta E\left( m\right) =
\frac V{2\pi ^2}\frac 12\int_0^\infty dpp_{}^3\left[ \sqrt{1-\left( \frac cp%
\right) _{}^2}+\sqrt{1+\left( \frac cp\right) _{}^2}-2\right] \text{ } \\  
\\ 
\text{becomes for }p_{}^2>>c_{}^2 \\  \\ 
\sim 
\frac V{2\pi ^2}\frac 12\int_0^\infty dpp_{}^3\left[ 1-\frac 12\left( \frac c%
p\right) _{}^2-\frac 18\left( \frac cp\right) _{}^4+1+\frac 12\left( \frac cp%
\right) _{}^2-\frac 18\left( \frac cp\right) _{}^4-2\right]  \\ 
\end{array}
$$
\begin{equation}
\label{ff21}=-\frac V{2\pi ^2}\frac{c_{}^4}8\int_0^\infty \frac{dp}p
\end{equation}
Introducing a cut-off one gets for the $UV$ limit\footnote{%
It is known that at one-loop level Gravity is renormalizable only in flat
space. In a dimensional regularization scheme its contribution to the action
is, on shell, proportional to the Euler character of the manifold that is
nonzero for the Schwarzschild instanton. Although in our approach we are
working with sections of the original manifold to deal with these
divergences one must introduce a regulator that indeed appears in the
contribution of the energy density.} 
\begin{equation}
\label{ff22}\int_0^\infty \frac{dp}p\sim \int_0^{\frac \Lambda c}\frac{dx}x%
\sim \ln \left( \frac \Lambda c\right) 
\end{equation}
and $\Delta E\left( m\right) $ for high momenta can be estimated by the
following expression 
\begin{equation}
\label{ff23}\Delta E\left( m\right) \sim -\frac V{2\pi ^2}\frac{c_{}^4}{16}%
\ln \left( \frac{\Lambda _{}^2}{c_{}^2}\right) =-\frac V{2\pi ^2}\left( 
\frac{3m}{r_0^3}\right) ^2\frac 1{16}\ln \left( \frac{r_0^3\Lambda _{}^2}{3m}%
\right) .
\end{equation}
At this point we can compute the total energy, namely the classical
contribution plus the quantum correction up to second order. Recalling the
definition of asymptotic energy for an asymptotically flat background, like
the Schwarzschild one 
\begin{equation}
\label{ff24}E_{{\cal ADM}}^{}=\lim _{r\rightarrow \infty }\int_{{\cal %
\partial M}}^{}\sqrt{\hat g}\hat g^{ij}\left[ \hat g_{ik,j}-\hat g%
_{ij,k}\right] dS^k,
\end{equation}
where $\hat g_{ij}$ is the metric induced on a spacelike hypersurface ${\cal %
\partial M}$ which has a boundary at infinity like $S^2$, one gets, 
\begin{equation}
\label{ff25}M-\frac V{2\pi ^2}\left( \frac{3m}{r_0^3}\right) ^2\frac 1{16}%
\ln \left( \frac{r_0^3\Lambda _{}^2}{3m}\right) =M-\frac V{2\pi ^2}\left( 
\frac{3MG}{r_0^3}\right) ^2\frac 1{16}\ln \left( \frac{r_0^3\Lambda _{}^2}{%
3MG}\right) 
\end{equation}
One can observe that 
\begin{equation}
\label{ff26}\Delta E\left( m\right) \rightarrow \infty \text{ when }%
m\rightarrow 0\text{, for }r_0=2m=2GM
\end{equation}
and 
\begin{equation}
\label{ff27}\Delta E\left( m\right) \rightarrow 0\text{ when }m\rightarrow 0%
\text{, for }r_0\neq 2m=2GM.
\end{equation}
{\bf Remark } We would like to explain the reasons that support the results
of formula $\left( \ref{ff23}\right) $. In that formula we introduced a
particular value of the radius, which behaves as a regulator with respect to
the horizon approach of the potential. The meaning of this particular value
is related to the necessity of explaining the dynamical origin of black hole
entropy by the entanglement entropy mechanism and by the so-called ``{\it %
brick wall model}'' \cite{t Hooft}. Indeed, the same mechanism is present
when one has to regularize entropy by imposing a kind of cut-off, that in
coordinate space means $r_0^{}>2m.$

\section{Summary and Conclusions}

\label{p6}

The trial wave functional approach, by means of Gaussian configurations, led
to possible calculations of quantum fluctuations of the gravitational field
around some fixed background geometry. In particular we have studied a
spherically symmetric background and by means of Birkhoff 's theorem we can
claim that our background is of the Schwarzschild type. Since we have
performed this analysis without any matter contribution and recalling the
definition of Ref. \cite{MTW} our result is valid for a Schwarzschild
wormhole. However this calculation apparatus is entirely based on the
possibility of explicitly breaking the invariance under reparametrisations,
expressed by the gauge fixing $\left( \ref{a4}\right) $, leading to the
conclusion that the final result seems depending on the foliation we choose
to work. For this reason to restore the invariance under reparametrisations,
we need to sum on every lapse function\footnote{%
A detailed version of this procedure will be studied in a future paper \cite
{Remo}}. With the gauge choice $\left( \ref{a4}\right) $, the problem of
defining a correct vacuum energy on every slice is well posed and the result
shows us an intrinsic energy depending only on the dynamics generated by
3-surfaces.

\section{Acknowledgments}

I wish to thank G. Esposito, V.Frolov, E. Gozzi, R. Parentani, D.L.
Rapoport, E. Recami for helpful discussion and P. Saurgnani who gave me the
technical support for the realization of this work. I also thank S. Liberati
and B. Jensen who suggested me how to justify the horizon approach.

\appendix

\section{Conventions}

Here we give the conventions for the metric tensor, connections and the
curvature tensor:

\begin{enumerate}
\item  Background Metric 
\begin{equation}
\label{ap1}
\begin{array}{c}
g_{11}=
\frac 1{1-\frac{2m}r},\text{ }g_{22}=r^2,\text{ }g_{33}^{}=r^2\sin ^2\theta 
\\  \\ 
g_{}^{11}=1-\frac{2m}r,\text{ }g_{}^{22}=r^{-2},\text{ }g_{}^{33}=r^{-2}\sin
^{-2}\theta 
\end{array}
\end{equation}

\item  Connection 
\begin{equation}
\label{ap2}
\begin{array}{c}
\Gamma _{ab}^1=\left( 
\begin{array}{ccc}
-\frac m{r^2}\left( 1-\frac{2m}r\right) ^{-1} &  &  \\  
& -\left( 1-\frac{2m}r\right) r &  \\  
&  & -\left( 1-\frac{2m}r\right) r\sin _{}^2\theta 
\end{array}
\right)  \\  
\\ 
\Gamma _{ab}^2=\left( 
\begin{array}{ccc}
0 & r^{-1} & 0 \\ 
r^{-1} & 0 & 0 \\ 
0 & 0 & -\sin \theta \cos \theta 
\end{array}
\right) \text{ }\Gamma _{ab}^3=\left( 
\begin{array}{ccc}
0 & 0 & r^{-1} \\ 
0 & 0 & \cot \theta  \\ 
r^{-1} & \cot \theta  & 0
\end{array}
\right) \text{ }
\end{array}
\end{equation}

\item  Riemann tensor, Ricci tensor and the Scalar Curvature in $3D$%
$$
R_{ijm}^l=\Gamma _{mi,j}^l-\Gamma _{ji,m}^l+\Gamma _{ja}^l\Gamma
_{mi}^a-\Gamma _{ma}^l\Gamma _{ji}^a\text{ \ Riemann tensor} 
$$
Because of the vanishing of the Weyl tensor in $3D$, that is $C_{ijm}^l=0$,
Riemann tensor is completely determined by Ricci tensor%
$$
R_{lijm}^{}=g_{lj}R_{im}^{}-g_{lm}R_{ij}^{}-g_{ij}R_{lm}^{}+g_{im}R_{lj}^{} 
$$
$$
R_{im}=R_{ilm}^l\text{ \ Ricci tensor} 
$$
$$
R=g_{}^{lj}R_{lj}^{}\text{ \ Scalar curvature} 
$$
\end{enumerate}

\section{Scalar Curvature Expansion}

In this part we give the necessary tools for the scalar curvature expansion
in terms of the fluctuations of 3-surfaces around the background geometry.
Metric tensor will be separated in a classical part (background) plus a
quantum part i.e.

\begin{equation}
\label{app1}g_{ij}=\bar g_{ij}+h_{ij} 
\end{equation}

\begin{enumerate}
\item  Expansion of the determinant 
\begin{equation}
\label{app2}
\begin{array}{c}
\sqrt{g_{ij}}=\exp Tr\ln \sqrt{g_{ij}}=\exp Tr\frac 12\ln \left( \bar g%
_{ij}+h_{ij}\right) =\exp Tr\frac 12\left[ \ln \bar g_{ij}\left( 1+\frac{%
h_{ij}}{\bar g_{ij}}\right) \right] = \\  \\ 
\exp Tr
\frac 12\left[ \ln \bar g_{ij}+\ln \left( 1+\frac{h_{ij}}{\bar g_{ij}}%
\right) \right] \simeq  \\  \\ 
\sqrt{\bar g_{ij}}\left( 1+\frac 12h-\frac 14h_k^ih_i^k+\frac 16%
h_k^ih_l^kh_i^l-\frac 18hh_k^ih_i^k+\frac 18h^2+\frac 18h^3\right) +O\left(
h^4\right) 
\end{array}
\end{equation}

\item  The inverse metric expansion 
\begin{equation}
\label{app3}g^{ij}=\frac{\bar g_k^i}{\delta _k^j+h_k^j}\simeq \bar g%
^{ij}-h^{ij}+h^{ik}h_k^j-h^{ik}h_k^lh_l^j+h^{ik}h_k^lh_l^mh_m^j+O\left(
h^5\right) 
\end{equation}

\item  Connection%
$$
\Gamma _{ij}^k\stackrel{}{:=}\frac 12g^{kl}\left(
g_{li,j}+g_{lj,i}-g_{ij,l}\right)  
$$
To $0^{th}$ order $\Gamma _{ij}^k\stackrel{}{=\Gamma _{ij}^{k\left( 0\right)
}\stackrel{}{=}\frac 12\bar g^{kl}\left( \bar g_{li,j}+\bar g_{lj,i}-\bar g%
_{ij,l}\right) }.$

To $1^{st}$ order $\Gamma _{ij}^k\stackrel{}{=S_{ij}^k\stackrel{}{=}\frac 12%
\bar g^{kl}\left( h_{li|j}+h_{lj|i}-h_{ij|l}\right) }$

The higher order corrections to the connection are related by the formula 
\begin{equation}
\label{app4}\Gamma _{ij}^{k\left( n\right) }\stackrel{}{=-h_l^k\Gamma
_{ij}^{l\left( n-1\right) }\text{ where }\Gamma _{ij}^{k\left( 1\right)
}=S_{ij}^k}
\end{equation}

\item  Riemann Tensor%
$$
R_{ijm}^l=\Gamma _{mi,j}^l-\Gamma _{ji,m}^l+\Gamma _{ja}^l\Gamma
_{mi}^a-\Gamma _{ma}^l\Gamma _{ji}^a 
$$
It is convenient to divide Riemann tensor into two terms: linear and
non-linear 
\begin{equation}
\label{app5}
\begin{array}{c}
{\it Lin}\left( R_{ijm}^l\right) =L_{ijm}^l:=\Gamma _{mi,j}^l-\Gamma
_{ji,m}^l \\  \\ 
{\it N-Lin}\left( R_{ijm}^l\right) =N_{ijm}^l:=\Gamma _{ja}^l\Gamma
_{mi}^a-\Gamma _{ma}^l\Gamma _{ji}^a
\end{array}
\end{equation}
The higher order terms of $L_{ijm}^l$ are simply 
\begin{equation}
\label{app6}
\begin{array}{c}
L_{ijm}^{l\left( n\right) }:=\Gamma _{mi|j}^{l\left( n\right) }-\Gamma
_{ji|m}^{l\left( n\right) } \\ 
\end{array}
\end{equation}
while higher orders of $N_{ijm}^l$ are 
\begin{equation}
\label{app7}N_{ijm}^{l\left( n\right) }:=\sum_{j=1}^{n-1}\left[ \Gamma
_{mi}^{a\left( j\right) }\Gamma _{ja}^{l\left( n-j\right) }-\Gamma
_{ma}^{l\left( n-j\right) }\Gamma _{ji}^{a\left( j\right) }\right] 
\end{equation}

\item  Second order scalar curvature
\end{enumerate}

Collecting together previous expansion formulas we can write the following
expression for the $^3R$ scalar curvature expanded up to second order: 
\begin{equation}
\label{app8}\int d_{}^3x\left[ -\frac 14h\triangle h+\frac 14h^{li}\triangle
h_{li}-\frac 12h^{ij}\nabla _l\nabla _ih_j^l+\frac 12h\nabla _l\nabla
_ih_{}^{li}-\frac 12h^{ij}R_{ia}h_j^a+\frac 12hR_{ij}h_{}^{ij}\right] . 
\end{equation}


\begin{references}
\bibitem{Garay}  L. J. Garay, Phys. Rev. D {\bf 48, }1710 (1993).

\bibitem{MazMot}  P. O. Mazur and E. Mottola, Nucl. Phys. {\bf B 341}, 187
(1990), D. Giulini, Phys. Rev. D {\bf 10} 5630 (1995).

\bibitem{York}  J. W. York Jr., J. Math. Phys., {\bf 14}, 4 (1973).

\bibitem{ADM}  R. Arnowitt, S. Deser, and C. W. Misner, in {\it Gravitation:
An Introduction to Current Research,} edited by L. Witten (John Wiley \&
Sons, Inc., New York, 1962); B. S. DeWitt, Phys. Rev. {\bf 160}, 1113 (1967).

\bibitem{Kerman}  A. K. Kerman and D. Vautherin, Ann. Phys., {\bf 192}, 408
(1989); J. M. Cornwall, R. Jackiw and E. Tomboulis, Phys. Rev. D {\bf 8},
2428 (1974); R. Jackiw in {\it S\'eminaire de Math\'ematiques Sup\'erieures,
Montr\'eal, Qu\'ebec, Canada- June 1988 - Notes by P. de Sousa Gerbert}; M.
Consoli and G. Preparata, Phys. Lett. B, {\bf 154}, 411 (1985).

\bibitem{t Hooft}  G.`t Hooft, Nucl. Phys. B{\bf \ 256} (1985), 727; V.P.
Frolov, I.Novikov, Phys. Rev. D {\bf 48 }(1993), 4545.

\bibitem{Regge}  T. Regge and J. A. Wheeler, Phys. Rev. {\bf 108}, 1063
(1957)

\bibitem{Miller}  A. Kheyfets and W. A. Miller, gr-qc/9412037

\bibitem{J.J. Halliwell}  J.J. Halliwell, ``Introductory Lectures on Quantum
Cosmology''. In {\it Jerusalem Winter School for Theoretical Physics:
Quantum Cosmology and Baby Universes Vol. 7}. S.Coleman, J.B. Hartle, T.
Piran and S. Weinberg, eds. World Scientific, 159-243.

\bibitem{MTW}  C.W.Misner, K.S. Thorne and J.A. Wheeler, {\it Gravitation}
(Freeman, San Francisco, 1973) 842; M.S. Morris and K.S. Thorne, Am. J. Phys.%
{\it \ }{\bf 56 }(1988) 395.

\bibitem{Remo}  R. Garattini, in preparation.
\end{references}
\end{document}